\documentstyle[12pt,aaspp4]{article}

\def\ni{\noindent}

\def\bar{\overline}

\def\bbE{\bar {\bf E}}

\def\beq{\begin{equation}}
\def\ee{\end{equation}}
\def\lsim{\mathrel{\rlap{\lower4pt\hbox{\hskip1pt$\sim$}}
    \raise1pt\hbox{$<$}}}
\def\gsim{\mathrel{\rlap{\lower4pt\hbox{\hskip1pt$\sim$}}
    \raise1pt\hbox{$>$}}}

\def\bfi{{\bf i}}

\def\bB{\overline B}
\def\bA{\overline A}
\def\bV{\overline V}
\def\ts{\times}
\def\lb{\langle}
\def\rb{\rangle}
\def\curl{\nabla {\ts}}

\def\bbV{\bar {\bf V}}

\def\bfv{{\bf v}}

\def\bfb{{\bf b}}

\def\bbB{\overline {\bf B}}
\def\bbA{\overline {\bf A}}
\def\bbM{\overline {\bf M}}

\begin{document}
\title{Extracting Rotational Energy in Supernova Progenitors:
Transient Poynting Flux Growth vs. Turbulent Dissipation}

\author{Eric G. Blackman,\altaffilmark{1,2,3} Jason T. 
Nordhaus,\altaffilmark{1,2} John H. Thomas\altaffilmark{1,3,4}} 
\affil{1. Dept. of Physics and Astronomy, Univ. of Rochester,
    Rochester, NY 14627, USA; 2. Laboratory for Laser Energetics, Univ. of Rochester, 
    Rochester, NY 14623, USA; 3. Isaac Newton Institute for Mathematical Sciences, 
    Univ. of Cambridge, Cambridge CB3 0EH; 4. Dept. of Mechanical Engineering, Univ. of 
    Rochester, Rochester, NY 14627, USA}
\centerline{(submitted to New Astronomy)}

\begin{abstract}
Observational evidence for anisotropy in supernovae (SN) 
may signal the importance of angular momentum and differential rotation in 
the progenitors.  
Free energy in differential rotation and rotation can be extracted 
magnetically or via turbulent dissipation. 
The importance that magnetohydrodyamic jets and coronae may play in driving SN
motivates understanding large scale dynamos in SN
progenitors.  We develop a dynamical large scale interface dynamo model 
in which the differential rotation and rotation deplete both through
Poynting flux  and turbulent diffusion.
We apply the model to a differentially rotating core surrounded by a
convection zone of a SN progenitor from a initial 15$M_\odot$ star.
Unlike the Sun, the dynamo is transient because 
the differential rotation  is primarily due to the initial collapse.
Up to $\sim 10^{51}$erg can be drained
into time-integrated Poynting flux and heat, the relative
fraction of which depends on 
the relative amount of turbulence in the shear layer vs. convection zone
and the fraction of the shear layer into which the magnetic field penetrates.
Both sinks can help facilitate explosions
 and could lead to different levels of anisotropy and pulsar kicks.  
In all cases, the  poloidal magnetic field is much weaker than the toroidal
field, and the  Poynting flux is lower than previous 
estimates which  invoke the magnitude of the total magnetic energy. 
A signature of  a large scale
dynamo is that the oscillation of the associated
 Poynting flux on $\sim 1$ sec time scales, implying the same for the  energy delivery to a SN.

\end{abstract}

\ni {\bf Key Words}: {supernovae: general -- stars: magnetic fields -- stars: neutron -- MHD -- dynamo theory--Gamma rays: bursts}

\section{Introduction}
Observations suggest that many supernovae (SN) are intrinsically
anisotropic.  As summarized by 
 Wheeler (2004a,b) evidence
comes from: (1) bipolar structure in supernova remnants
(Dubner et al.\ 2002); 
(2) optical jet and counter jet structures in Cas A (Fesen 2001);
(3) X-ray jets and toroidal emission of intermediate mass elements (Hughes
 et al.\ 2000; Hwang, Holt, Petre 2000; Willingale et al.\ 2002);
(4) asymmetric ejecta in SN1987A perpendicular to the major axis of the
rings (
(Wang et al.\ 2001, 2002); and
(5) Optical polarization in both Type I and Type II  
supernovae (see Wang 2004 for a review)
with the Type Ib,c polarization being higher than that
of Type II (Wang et al. 1996; Wang et al.\ 2001), 
but with the latter increasing with time (Wang et al. 2001; 
Leonard et al. 2001). Since Type I SN represent a naked core, and the core at late times in Type II 
becomes exposed, a consistent interpretation is that the  
SN engine is a source of this asymmetry.
At minimum, anisotropy likely reveals the role of rotation in SN engines and the need to incorporate non-spherically symmetric physics.  

\subsection{Magnetic Fields}
The combination of rotation, highly ionized plasma, stratified  
turbulence, and magnetic fields has prompted various considerations of MHD outflows 
as a source of this asymmetry, if not the source of the SN explosion itself.
Recent renewed interest in  proposals relating MHD outflows 
to supernovae (Leblanc \& Wilson 1970; Meier et al.\ 1976; 
Ardeljan, Bisnovatyi-Kogan, \& Moiseenko 1998; Khokhlov et al.\ 1999; 
Wheeler et al.\ 2000; Wheeler, Meier, \& Wilson 2002; 
Moiseenko, Bisnovatyi-Kogan, \& Ardeljan 2004) is bolstered by the  
observational association of SN with Gamma-ray bursts 
(Galama et al 1998; Iwamoto et al.\ 1998; 
Stanek et al. 2003; Hjorth et al. 2003; Thomsen et al. 2004; Cobb et al.\ 2004
Gal-Yam et al. 2004; Malesani et al. 2004).
MacFadyen, Woosley, \& Heger 2001).
That magnetized outflows have also long been thought to be important in young 
stellar objects, active galactic nuclei, pulsar winds, and  microquasars 
(of which GRB are likely one class) suggests  that explosive bipolar outflows 
in astrophysics may 
all involve some combination of rotation and magnetic fields. 

Previous work on  magnetic field amplification in pre-supernovae
core conditions have employed either (i) simple shear that grows
toroidal field linearly (e.g. Leblanc \& Wilson 1970; 
Ardeljan, Bisnovatyi-Kogan, \& Moiseenko 1998; Wheeler 2000), 
(ii) traditional kinematic convective $\alpha-\Omega$ dynamo models to grow 
ordered fields (Thompson \& Duncan 1993), which in principle grow large-scale 
fields exponentially, or (iii) the exponential field growth from the 
magneto-rotational instability (MRI, e.g. Balbus \& Hawley 1998) 
(Akiyama et al.\ 2003; Moiseenko et al.\ 2004),
whose saturated magnetic energy was used as a rough estimate for the 
large-scale toroidal field. Each has its merits and its limitations.
In the presence of turbulent diffusion, (i) will not sustain the field. 
The approach of (ii) offers exponential growth and is an important step 
forward, but the backreaction of the field on the driving flow, the separate 
generation of toroidal vs. poloidal fields, and the spatial location of the 
$\alpha$ and $\Omega$ effects still need to be determined. 
The recent approach of Akiyama et al.\ in (iii) offers a useful model of the 
internal and rotational structure of the inner SN engine, 
and also makes careful estimates of the saturated energy of the magnetic
field adopted from results obtained in MRI disk simulations.
Whether these values also apply to pressure-supported stars remains to be 
studied in detail because the MRI likely 
transports angular momentum on spherical
shells (Balbus \& Hawley 1994), not radially; the latitudinal
differential rotation may be most important.
If 3-D turbulence develops, then field amplification will take place from the MRI. 
Moiseenko et al.\ (2004) provide 2-D simulations of what appear to be MRI-driven SN, 
starting from a relatively strong, ordered poloidal field. 
However, in 2-D, there is not
sustained dynamo action to amplify the total magnetic energy from 
arbitrarily small values.

An important question is whether a magnetically driven outflow 
requires a large-scale field in the supernova progenitor engine.
By large-scale field, we mean one that maintains the sign of its flux over many 
dynamical times, when averaged over the size of the engine region. 
Large-scale fields are helpful, if not
necessary, in  generating outflows in disks or stellar coronae, and perhaps by analogy, also in SN engines. 
If the field needs to  buoyantly rise
above the dynamo region to a height at which it provides the 
dominant contribution to the stress, 
the field should be of large enough scale to 
avoid being shredded by turbulence. Once at the region where
it dominates, the field can further relax to even larger scales. 
Comparing the shapes of observed outflows to those from theoretical studies
(e.g. Moiseenko et al.\ 2003) may help determine the relative importance of
large scale fields vs. a simple magnetic pressure gradient.

This magnetic field may drive a  ``jack-in-the-box''
type of  magnetic spring explosion.  
(In addition to Wheeler, Meier, \& Wilson (2002), 
see also Matt et al.\ (2004) for simulations of a magnetic explosion without 
specifying the field origin, and Moiseenko et al.\ (2004) for a magnetic explosion 
driven SN.).
%
The potential efficacy of magnetically-driven outflows for SN is evident from recent work by
Moiseenko et al.\ (2004). However, 
as in the magnetic explosion of Matt et al (2004) 
for planetary nebulae, Moiseenko et al. (2004) 
start with a significant large-scale poloidal 
field, comparable in strength to the saturated poloidal field 
whose growth we will study 
here from an initially small seed field.  
Any dynamo driven outflow must occur while the 
dynamo is active because afterward the field decays.

Large-scale field generation does not exclude the presence of the MRI:
When the MRI operates in a stratified medium, it may provide 
a source of helical turbulence that allows one sign of magnetic helicity
to migrate to large scales, producing large-scale magnetic fields
(e.g. Brandenburg et al.\ 1995; Blackman \& Tan 2004)
or a helicity flux (Blackman \& Field 2000; Vishniac \& Cho 2001).
In our large scale dynamo herein, the source of helical turbulence is considered to be
convection driven by shock heating and neutrino 
deposition rather than the MRI, though the particular source is not essential
for our calculations.
It should also be emphasized that  studies focusing on magnetic energy amplification 
from the MRI (Balbus \& Hawley 1998) typically study growth of 
the total magnetic energy 
(e.g. Akiyama et al. 2003) rather than a specifically 
large-scale  field in the sense we have
defined above.  
The backreaction of the field 
on the shear is not usually considered in MRI studies. 

Note that instead of a bulk dynamical influence of large scale fields, 
Ramirez-Ruiz and Socrates (2004) highlight an alternative role 
facilitated by large scale fields that buoyantly rise
to form a corona. They  suggest that such a corona could 
distribute the binding energy 
dissipation such that the neutrino spectrum becomes non-thermal.
This increases the efficiency of neutrino-matter coupling by an
order of magnitude. The SN would be neutrino driven, but 
symbiotically dependent on a corona.

\subsection{Turbulent Viscosity}

Magnetic fields generated in situ are amplified by extraction of
the free energy in differential rotation. 
Thompson et al. (2001) point out that the energy lost into heat
via turbulent damping of the shear supplies enough energy when
combined with neutrino driving to create a SN explosion. 
The energy lost from the shear 
to heat represents a conversion of some of the binding energy into a form
more efficiently coupled to the matter than neutrinos.
In the presence of turbulence (which itself could be induced by
the magnetic field) a competition emerges between
draining the shear into Poynting flux  vs. heat.
Both sinks may help drive  SN in different ways.

\subsection{What We Do Here}

Here we study the dynamical evolution of differential
rotation and large scale magnetic field growth 
with several key new features: 1) we explicitly consider the backreaction
of the magnetic field on the shear and rotation driving the field growth and
2) allow the shear and rotation to deplete via Poynting flux
and  turbulent dissipation.  The latter turns out
to be important and a key result of our study is to show
the relative  depletion into each channel.

We derive a dynamical generalization of the large-scale interface dynamo proposed by Parker (1993) for the Sun (see also Charbonneau \& MacGregor 1997; 
Markiel \& Thomas 1999; Zhang et al.\ 2003).
Interface dynamos differ  from the original $\alpha-\Omega$ 
dynamos in that the 
 dominant region of shear and the dominant region of helical 
turbulence are not co-spatial.
The shear associated with differential rotation is concentrated in a layer
(the $\Omega$-layer) lying beneath the turbulent convection zone (the $\alpha$ layer).
Such a scenario has been a leading model for the solar dynamo,
and has also been applied to white dwarfs (Markiel, Thomas \& Van Horn 1994; 
Thomas, Markiel \& Van Horn 1995) and to AGB stars (Blackman et al.\ 2001).  
In all of these cases, 
helical turbulence in the convective zone provides the dynamo
$\alpha$  effect and thus generates the poloidal field. Turbulent pumping
(e.g. Tobias et al.\ 2001) 
pushes the poloidal field downward into the strong shear layer where the toroidal 
field is amplified. The rotational and convective environment 
of the proto-supernova engine in core-collapse supernovae
has a similar structure, with a strong shear layer surrounded by
a convective envelope. However, unlike the 
sun where the differential rotation profile is re-established   
by convection, in the SN context the intial 
shear decays dynamically as the field grows. Whether or not 
convection may refuel the differential rotation needs
further study, but here we assume that the dominant source of
differential rotation is that from the initial collapse.
This differential rotation will deplete as the field is amplified
or as the shear is damped by turbulence.
We explicitly include the dynamical evolution of the shear and rotation
in addition to that of the magnetic field, unlike previous
work.  The  dynamo becomes transient in the SN progenitor.

Dynamo research must accommodate two competing needs: 
the need for a theory that rigorously includes the backreaction of the growing 
magnetic field on the flow, and the need to model the field growth in a realistic system.
Meeting these two goals is difficult both analytically and numerically, and  
compromises must be made on both fronts. 
Our aim here is to develop a  transient large scale dynamo 
model and apply it to the SN progenitor context as simply as possible, while 
still including the time-dependent nonlinear dynamical quenching
of the shear and  rotation.

In section 2 we present the dynamical equations for the magnetic field, 
shear, and rotation.
In section 3 we discuss the parameter choices and present the corresponding solutions.
In section 4 we discuss the implications of the calculated Poynting flux
and the fractional shear energy drained into magnetic energy vs. heat.
We conclude in section 5 and identify some unresolved issues for future work.

\section{Transient Interface Dynamo and Velocity Equations}
We generalize the  interface dynamo model of Markiel, Thomas \& Van Horn (1994) 
to include the dynamical evolution of shear, 
rotation,  and Poynting flux.  
The concept of the interface dynamo  is illustrated in the meridional 
slice shown in Fig. 1. 
Two distinct regions adjoin at the interface radius $r=r_c$. 
In the inner region, the 
differential rotation (the $\Omega$ effect) wraps the poloidal field 
into a toroidal field with the rotation profile initially varying linearly from $\Omega$ at 
$r=r_c$ to $\Omega+\Delta\Omega$ at $r=r_c-L$,
where $\Delta\Omega$ measures the differential rotation.
The values of the quantities at the initial time of the calculation
are indicated by  a subscript $0$, that is 
$\Omega_0,$ and $\Delta\Omega_0$.
The outer layer, representing the 
convection zone, rotates rigidly with angular velocity $\Omega_0$.
There, cyclonic convection (the $\alpha$-effect) converts toroidal field into 
poloidal field. This poloidal field is then pumped or diffused downward into the
$\Omega$ layer where the shear again further amplifies the toroidal field. 
The relevant $\alpha$ layer has thickness $L_c$, which
corresponds to a density scale height from the base of the convection zone.
A full treatment of the interface dynamo would include anisotropic diffusion,
stratification and shear, global 3-D treatment of the convection, 
self-consistent
 buoyancy dynamics,  self-consistent sustenance of $\alpha$  
and a dynamical treatment of the non-linear backreaction of 
magnetic field on the velocity field. 
We provide a minimalistic approach that incoporates some
of these ingredients in a 1-D model to illustrate the mechanism simply.

\subsection{Magnetic Field Evolution}
To solve for the magnetic field, we employ the mean-field induction 
equation, derived by averaging the induction equation in the presence of helical  
velocity fluctuations.  
The standard mean field induction equation, ignoring cross-helicity, is 
(e.g. Parker 1979)
\beq
\partial_t\bbB = \curl\lb\bfv\times\bfb\rb+\curl (\bbV\ts\bbB)
+\lambda\nabla^2\bbB
=\curl (\alpha \bbB) + \curl (\bbV\ts\bbB)-\curl (\beta_{\bfi} \curl \bbB)
+\lambda\nabla^2\bbB,
\ee
where $\bbB$ is the mean field, $\bbV$ is the mean velocity,
$\lambda$ is the turbulent diffusivity, 
and the turbulent electromotive force is $\lb\bfv\times\bfb\rb
= \alpha\bbB-\beta_{\bfi}\curl\bbB$, 
where $\alpha$ is the pseudoscalar helicity dynamo coefficient
and  $\beta_i$ is a turbulent magnetic diffusivity,
where the index allows different values for the poloidal and toroidal
field equations.
Although  we have in mind a spherical system, for present purposes we simplify 
the analysis by working in local Cartesian coordinates.  
Simplified versions of the equations we invoke  
for the mean magnetic field have been used in 
previous interface dynamo models (\cite{RD,MTVH}) without adequate derivations, and so we present a full derivation here.
The coordinates are shown in Fig. 1. 
Writing $\bbA=(\bA_x,\bA,\bA_z)$ and assuming that
(i) the mean fields are
axisymmetric (i.e. $\partial_y \bbM=0$ for any mean quantity $\bbM$),
(ii) $\bbV=(0,\bV,u)$, 
(iii) $\bbB=(0,\bB,\partial_x \bA)$ (so $\partial_z \bA=0$) 
(iv)  $\beta_{\bfi}=\beta_p$ for the poloidal field equation and
$\beta_{\bfi}=\beta_t$ for the poloidal field equation and
$\nabla \beta_t= \nabla \beta_p =0$. 
(v) $\beta_{\bfi}>>\lambda$  
we can write the equations for the toroidal 
magnetic field $\bB$ and the vector potential 
$\bA$ as
\beq
\partial_t\bB = -\alpha\partial_x^2\bA-\partial_z\alpha\partial_xA
+ \partial_x \bA\partial_z\bV -u\partial_z \bB-u\bB/L  +\beta_t\nabla^2 \bB
\label{6}
\ee
\beq
\partial_t\bA = \alpha\bB+\beta_p\partial_x^2 \bA,
\label{7}
\ee
where we have used $-\bB\partial_z u 
\sim -u\bB/L$ 
which we take as a loss term due to magnetic buoyancy with $u>0$.
If we assume that the Fourier transform of the field 
is proportional to a $\delta$ function in wavenumber
(i.e.  the mean field has a single large scale), then we can write 
$\bA=A(t)e^{(ik_x x+ik_z z)}$ and $\bB=B(t)e^{(ik_x x+ik_z z)}$, 
where $A(t)$ and $B(t)$ are complex-valued functions.
Then, assuming the $\partial_z\alpha$ term to be small,
Eqs. (\ref{6})  and (\ref{7}) become

\beq
\partial_tB 
\simeq \alpha k_x^2A 
-ik_x r_c A{\Delta \Omega\over L} -iuk_zB -uB/L-\beta_t k^2 B 
\label{8}
\ee
\beq
\partial_tA = \alpha B-\beta_pk^2  A,
\label{9}
\ee
 where $k^2=k_x^2 +k_z^2$, and we have used 
\beq
\partial_z\bV \sim -r_c \Delta \Omega/L
\label{9a}
\ee
to obtain the third term in (\ref{8}), based
on the interface geometry shown in Fig. 1.
We write
\begin{equation}
\beta_t=c_t{v}L  \ \ \ {\rm and}\ \ \ 
\beta_p=c_p{v}L 
\label{diffparam}
\end{equation}
where $c_t,c_p$ are distinct
dimensionless constants that allow our 1-D model
to account for the fact that toroidal and poloidal fields are generated in separate regions with distinct turbulent diffusivities.
The quantity $v_1$ 
 is a typical convective velocity at the middle of the $\alpha$ layer (i.e. at $r=r_c+L_1/2$, see Fig. 1).  

In interpreting $-\frac{uB}{L}$ as the rate of loss of toroidal flux due to the buoyant rise of toroidal flux out of the dynamo region, 
we use the expression for the rise velocity of a flux tube (Parker 1955, 1979), namely
\begin{equation}
u=\frac{3Q}{8}{\left(\frac{a}{L}\right)}^2\frac{|\bB|^2}{4\pi\rho{v}}=\frac{3Q}{32}\frac{\bV_A^2}{v},
\label{buoy}
\end{equation}
where $a$ is the radius of the flux tube (assumed to be $L/2$), $\bV_A$ is 
the Alfv\'{e}n speed associated with $\bB$, 
and $Q$ is a dimensionless constant.  
\subsection{Quenching of $\alpha$}
While there is no compelling evidence that $\beta$ is catastrophically quenched in convective 3-D MHD 
turbulence, recent work on dynamo theory has shown that $\alpha$ quenching can be 
understood dynamically via magnetic helicity conservation
(e.g. Blackman \& Field 2002;
Brandenburg \& Subramanian 2004 for a review).
The build-up of the large-scale field is associated with a build-up of the 
large-scale magnetic helicity, $\bbA\cdot\bbB$.
In the absence of boundary terms, magnetic helicity is well conserved,
and the small-scale helicity builds up to equal magnitude but
opposite in sign  to that of the large scale. Since $\alpha$ depends on the difference between kinetic and
current helicities, the build up of small-scale magnetic (and thus current) 
helicity eventually saturates the dynamo. If boundary terms 
can allow for a helicity flux 
(Blackman \& Field 2000; Vishniac and Cho 2001) 
then the $\alpha$ quenching is alleviated but at the expense of a lower
saturated mean field 
(Blackman \& Brandenburg 2003). 
If the small-scale helicity can be preferentially
removed then the large-scale growth proceeds longer to larger values.
Shear may also reduce the impact of 
$\alpha$ quenching (Brandenburg \& Sandin 2004).
Shear also leads to anisotropic turbulence and possibly a shear-current
dynamo (Rogachevskii \& Kleeorin 2003), 
though  we do not  presently discuss this further.

Here we are interested in demonstrating the most basic aspects
of a  transient large scale dynamo and adopt a parameterization of the 
$\alpha$ effect backreaction that can be used to approximate 
the non-linear quenching.
We modify 
the quenching formula used for previous interface dynamos (\cite{MTVH}) 
to also facilitate the coupling of $\Omega$ quenching into $\alpha$.
We write
\begin{equation}
\alpha=\alpha_0 (\Omega/\Omega_0)^q
{Exp}\left[-\gamma_1{\frac{\bB^2/8\pi}{\rho_1v_1^2/2}}\right],
\label{8aa}
\end{equation}
where $\gamma_1$ is a dimensionless constant,  
$\rho_1$ and $v_1$ are 
the mass density and a typical convective velocity at the middle of the 
$\alpha$ layer.
We assume $q=1$ which so that $\alpha$ quenching   
is coupled  to the $\Omega$ quenching discussed
below,  and use (\cite{DR})
\begin{equation}
\alpha_0=c_\alpha\frac{L_1^2\Omega_0}{r_c},
\end{equation}
where $0< c_\alpha< 1$ is a dimensionless constant.  


\subsection{Evolution of  $\Delta \Omega$ and $\Omega$}
The amplification of the field and turbulent diffusion 
will drain the differential rotation  and Poynting flux  
drains some of the remaining rotational energy until the dynamo shuts off.
We must therefore construct dynamical equations for $\Delta \Omega$ and $\Omega$.


If we ignore second derivatives in space,  
 using (\ref{9a}) we can write the differential equation
for the  shear at $r=r_c$ as 
\beq
-\partial_t(r_c\Delta \Omega) =
\partial_t (\bV_y(r_c)-\bV_y(r_c-L))\simeq\partial_t(L\partial_x \bV_y).
\label{14}
\ee
To obtain the desired differential equation for the right hand side,
we subtract the time dependent differential equations for the velocity at
$r_c$ and $r_c-L$.  From the Navier-Stokes equation for the mean velocity 
and the assumption that
gradients in the $y$ direction vanish, 
\beq
\partial_t\bV_y= -\bbV\cdot\nabla\bV_y + {1\over 4\pi \rho}(\bbB\cdot\nabla)\bB_y+\nu\nabla^2\bar{V}_y\simeq
{1\over 4\pi \rho}\partial_x\bA \partial_z\bB_y+\nu(\partial_x^2+\partial_z^2)V_y,
\label{15}
\ee
where $\nu$ is a turbulent viscosity. 
From Eqs. (\ref{14}) and (\ref{15}) we then have 
\beq
\begin{array}{r}
\partial_t(r_c\Delta\Omega)=
({1\over 4\pi \rho}\partial_x\bA \partial_z\bB_y+\nu(\partial_x^2+\partial_z^2)v_y)|_{r_c}-
({1\over 4\pi \rho}\partial_x\bA \partial_z\bB_y+\nu(\partial_x^2+\partial_z^2)v_y)|_{r_c-L}
\\
\simeq  {L\over 4\pi}\partial_z{1\over \rho}\partial_x\bA \partial_z\bB_y+\nu{L}(\partial_z\partial_x^2+\partial_z^3)v_y \\
\simeq{L\over 4\pi}\left[{1\over \rho}
(\partial_z\partial_x\bA \partial_z\bB_y)_{r_c}+
(\partial_x\bA \partial_z^2\bB_y)_{r_c}
-({\partial_z\rho\over \rho^2}\partial_x\bA \partial_z\bB_y)_{r_c}
\right]-\frac{\nu{r_c}}{L^2}\Delta\Omega.
\end{array}
\ee
Taking $\partial_z\rho \sim (\rho_2-\rho_1)/L$,  where $\rho_2$ and $\rho_1$
are the densities at the inner and outer boundaries of the shear layer, 
and using $\bA=A(t)e^{i(k_x x +k_z z)}$ 
and $\bB=B(t)e^{i(k_x x +k_z z)}$ as in the derivation of Eqs. (\ref{8}) and (\ref{9}), 
we then obtain 
\beq
\partial_t\Delta\Omega={L\over 4\pi r_c \rho}\left[-k_xk_z^2\rho (Re(\bA)Re(i\bB)+Re(i\bA)Re(\bB))-\partial_z \rho k_x k_z Re(i\bA)Re(i\bB)\right]-\frac{\nu}{L^2}\Delta\Omega.
\label{dom}
\ee

As stated above, we also need an equation for the time evolution of 
$\Omega(t)$. We obtain this equation by noting that the rotational
energy of the field-anchoring matter is drained by the 
Poynting flux.  The Poynting flux at $r_c$ is given by 
\begin{equation}
L_{mag}={c\over 4\pi}\int {\bbE}\times {\bbB} \cdot d{\bf S}_c \simeq 
{1\over 4\pi}\int \Omega r_c Re(\bB_z)Re({ \bB}_y) \cdot d{\bf S}_c 
\simeq -Re(\bB_z) Re(\bB_y) \Omega r_c^3.
\label{pf}
\end{equation}
Calculating this quantity  
requires the separate determination of the toroidal and poloidal magnetic 
fields: because these components can be out of phase, 
the maximum Poynting flux is not simply the product of their respective 
maxima. 
We approximate the total rotational energy in the shear layer as the 
kinetic energy of the
of mass $M\simeq 10^{34}$g of the shear layer.  
moving with the velocity of the outer boundary, namely, 
$E_{rot}\sim Mr_c^2{\Omega}^2/2$. This value is imprecise but
gives between $10^{51}\le  E< 10^{52}$, which is 
consistent with other estimates of the available rotational energy 
(Thompson et al. 2005).
However, because the fields may not penetrate into the entire shear layer,
we allow for the fact that only a fraction of the
rotational energy may be available for conversion into Poynting flux.  
We account for this reduced available rotational energy by multiplying $E_{rot}$ by $\delta/L$, where $\delta$ is the depth penetrated by the 
field into the shear layer, a quantity will derive in (\ref{delta}) below.  
In combination with Eq. (\ref{pf}), the time derivative of 
the rotational energy then leads  to the time-evolution equation for $\Omega$
\beq
\partial_t \Omega\simeq  {Re(\bB_z) Re(\bB_y) r_c\over M \delta}
\label{omdif}
\ee

Eqs. (\ref{8}), (\ref{9}), (\ref{dom}) and 
(\ref{omdif}), 
are the coupled differential equations to be solved for the transient
large scale interface dynamo.
In all solutions discussed below, we will take $\beta_t=\nu$,
and define the 
the poloidal turbulent magnetic Prandtl number 
\beq
P_{M,p}\equiv\frac{\nu}{\beta_p}=\frac{\beta_t}{\beta_p}.
\label{prandtl}
\ee
The role that $P_{M,p}<1$ plays is very important:
Although the equations we solve are 1-D, we invoke 
our use of $P_{M,p}<1$ is a simple way of capturing  
aspects of a 2-D interface dynamo. In particular, in a realistic
progenitor the shear
layer is convectively stable and thus can be expected to have
a lower turbulent diffusion coefficient than the convective
zone. In addition, the toroidal field is primarily amplified 
in the shear layer where as the poloidal field
is primarily amplified in the convection zone. 
We therefore  take $\beta_p>> \beta_t=\nu$.

\section{Discussion of Solutions}
\subsection{Kinematic Solution}
For the relevant parameters,
the first term on the right of (\ref{8}) is small compared to the
next term, implying that we are initially in the $\alpha-\Omega$ interface
dynamo regime.
It is useful to first consider 
the kinematic limit, $\gamma_1=q=0$, $\partial_t f=\partial_t\Omega=0$,
which allows us to determine the conditions for initial growth of the dynamo field.
In this case, 
(\ref{8}) and (\ref{9}) are the only equations to be solved.
Assuming that 
$k_z << k_x $, 
there are then exponentially growing solutions of the form $A(t)=A_0e^{nt}$
and $B(t)=B_0e^{nt}$,
such that $\bA=Re (A_0 e^{nt+ikx})$ and $\bB=Re (B_0 e^{nt+ikx})$,
and  
\begin{equation}
Im(n)=\left({\alpha_0\Delta \Omega k r_c \over 2L}\right)^{1/2}.  
\label{df}
\end{equation}
The amplitude of these waves grows (i.e. $Re(n)>0$) only when  the initial dynamo number
\begin{equation}
N_{D,0}\equiv N_D(0)=\frac{\alpha_0k r_c\Delta\Omega_0}{2L\beta_t\beta_pk^4}=\frac{\alpha_0k r_c\Delta\Omega_0}{2LP_{M,p}\beta_p^2k^4}
\label{dynnum}
\end{equation}
exceeds unity.  
 This kinematic solution will provide
insight into the non-linear solutions.

\subsection{Penetration Depth}

In our solutions  below, 
we will consider cases  in which 
the entire differential rotation layer is available for magnetic
field amplification  but also cases in which  
only a fraction $\delta/L$ of the layer is available.  
The estimate of $\delta$ for the latter case is obtained
by considering it to be the distance that the 
toroidal field can  diffuse into the shear layer
during a cycle period. The cycle period does increase
during the dynamical regime, but
as a lower limit  we use the kinematic value obtained 
from the inverse of (\ref{df}) namely $\tau=2\pi(\frac{2L}{\alpha_0\Delta\Omega_0kr_c})^\frac{1}{2}$. 
We then have 
\begin{equation}
\delta\simeq (\beta_t\tau)^\frac{1}{2}.
\label{delta}
\end{equation}

\subsection{Nonlinear Solutions for SN Progenitors}

Table 1 and Figs. 2-5 represent example solutions of 
Eqs. (\ref{8}), (\ref{9}), (\ref{dom}) and 
(\ref{omdif}). We discuss the choices of parameters and
the meaning of these solutions below.

For the  engine structure, we employ the 
profiles of Akiyama et al.\ (2003) for a rapidly rotating neutron star (NS)
formed from the collapse of a 15 $M_\odot$ progenitor. These authors used 
a 1-D stellar evolution code to obtain the radial structure of a spherical 
core collapse, and then computed the rotational velocity profiles 
a posteriori by assuming  that angular momentum is conserved on spherical
shells during the collapse.
We adopt the values of the density and rotation profiles
corresponding  to 384 milliseconds after bounce.  
While Akiyama et al.\ (2003) used their structure and rotation solutions 
as inputs for their MRI dynamo calculations, we use these
inputs for the interface dynamo of the previous section.

Using the characteristic numbers, our differential rotation layer ($\Omega$-layer, Fig. 1) 
extends down to the surface of the NS ($r_c-L=1.5\times{10^6}$ cm) and 
the base of the convection zone is located at $r_c=4\times{10^6}$ cm. 
We take the $\alpha$ effect to occur above the base of the convection
zone in a layer whose thickness ($L_1=2.3\times{10^6}$ cm)
equals the local density scale height.  The density in the middle of the $\Omega$ layer
is taken to be $\rho_2=2.4\times{10^{13}}$ g/cm$^3$ and 
that of the $\alpha$-layer is taken to be $\rho_1=2.4\times{10^{12}}$ g/cm$^3$ 
(Akiyama et al.\ 2003).  A typical convective velocity (e.g. Herant et al.\ 1994) 
of $v_1=10^8$ cm/s is taken in the middle of the $\alpha$ layer and 
is sustained by the neutrino luminosity and lepton gradients.  
The interface dynamo equations of the previous section contain 
the dimensionless constants $\gamma_1$, $c_t$, $c_p$, $c_\alpha$, and $Q$. 
For the $\alpha$ quenching parameter, $\gamma_1$, 
we take $\gamma_1=0.1$, for which the solutions
turn out to be $\Omega$-quenching limited.
Regarding the parameters  $c_t$, $c_p$, $c_\alpha$, and $Q$,
we note that the dynamo number is sensitive only to the ratio 
$c_\alpha/c_tc_p$, whereas $Q$ enters in the buoyancy loss term
and $c_\alpha$ enters in the kinematic dynamo frequency.
A range of choices can be considered to be appropriate given 
the uncertainty in the detailed properties of the turbulence, 
but, in the spirit of mixing-length theory, we choose values similar to those which yield dynamo solutions
matching the solar dynamo period and field strength when 
the theory (without $\Omega$ and $\Delta\Omega$ quenching) is applied in that 
context (\cite{MTVH}).

The magnetic field strengths $\bB_x, \bB_y$ are computed at the interface layer ($r=r_c$).
At $r=r_c$,  $x=r_c\theta$ locally where $\theta$ is the 
poloidal angle from the prescribed $z$ axis in Fig. 1.  
Although the Cartesian approach 
technically applies only locally, we choose the mode
that most closely corresponds to dipole symmetry, namely that for which $k_xr_c=2$.

Fig. 2 
shows the poloidal field, toroidal field,  
Poynting flux, $\Omega$, $\Delta \Omega$, and a comparison
of the time-integrated Poynting flux compared to the energy
in the differential rotation lost to turbulent dissipation 
for the parameter row 7 in Table 1. For Fig. 2 we
used (\ref{delta}) for $\delta$.
Note that the cycle period of the field growth is 
reflected in the oscillations of the Poynting
flux in the upper left panel.
The fact that the dotted curve is above the solid curve in the upper
right panel indicates that for this set of paramters, more of the free
energy is drained into
heat via turbulent dissipation rather than into Poynting flux.

Fig. 3 is similar to Fig. 2 but with parameters
used from row 4 in Table 1. The same $\beta_p\propto c_p$
is used as in Fig. 2 
but a lower $\beta_t\propto c_t$ is used and corresponding to  
a lower $Pr_{M,p}$.
From (\ref{dynnum}) we see that $N_D$ increases which implies 
shorter cycle period and  a shorter time for the field
to saturate compared to Fig. 2. A smaller $\beta_t$ and shorter
cycle period accounts for the lower $\delta/L$ compared
to Fig. 2, but a lower $\beta_t$ also implies less diffusion
in the equation for $B_\phi$ so $B_\phi$ peaks at a larger
value than in Fig. 2.

In Figs. 4 and 5 we show the cases of rows 17 and 18 in Table 1
respectively. For these cases
when computing the evolution of the total rotational energy,
we set $\delta=L$ to compare with Figs. 2-3 where (\ref{delta})
was used. In both of these cases, the total integrated Poynting
flux exceeds the energy lost in the shear layer via turbulent diffusion.
The difference between Figs 4 and 5 is the buoyancy
constant $Q$ which is smaller in Fig. 5 than in Fig. 4.

Table 1 shows a variety of other cases not shown in the figures,
but which can be analyzed similarly. One point to note, which is
evident from (\ref{dynnum}) is that the growth rate dependents
indepently on $Pr_{M,p}$ and $\beta_p$ (or equivalently, $c_p$ and $c_t$)
so that a fixing $Pr_{M,p}$ alone does not fix the saturation values.
This can be seen e.g. by comparing rows 4 and 5 of Table 1.

Taken collectively, the figures and Table 1 
illustrate the transient nature of the dynamo.
The transience and saturation are due to the depletion of
$\Delta \Omega$ and $\Omega$.
The former  depletes 
due to field amplification and turbulent diffusion 
and $\Omega$ depletes from the Poynting flux.
Most dynamo calculations in main sequence stellar contexts  and
accretion disks ignore $\Delta \Omega$ and $\Omega$ quenching and decay.
In the present SN progenitor context the dynamical evolution of these quantites
is important.  When $\Delta \Omega$ and $\Omega$ 
are enforced as constant, 
$\alpha$ quenching and buoyancy would  determine the maximum amplitudes of the
field and Poynting flux.  In that case,
the curves analagous to Figs. 2-5 would 
asymptote 
to oscillating curves of  constant peak amplitudes.

Increasing  $\Delta\Omega_0$ and $\Omega_0$ increases the initial
dynamo number $N_D$. This increases the rate of growth and 
maximum value of the fields obtained,
However, due to the finite amount of  
rotational energy, estimated above (\ref{omdif}), 
a faster growing dynamo lasts a shorter time.
This can also be seen by comparing Figs 2 and 3. The latter
has a faster growth to a higher peak, but decay from the peak is faster.
That 
$\Delta \Omega$ and $\Omega$ do not fall to $0$ before
the field decays in all of the figures results  because 
the effective dynamo number  becomes sub-critical well before all of the
rotational energy is drained.

The oscillation periods of the toroidal and poloidal fields 
are roughly consistent with  $2\pi/Im(n)$ 
at early times until the rise to the first maximum peak.
During this time, the Poynting flux oscillates with a period 
half that of the these field components.
After the field and Poynting flux rise to their absolute maxima, 
the cycle period increases slightly.  
This increase occurs because  
the cycle period in Eq. (\ref{df}) is 
$\propto \Omega^{-1/2}\Delta\Omega^{-1/2}$,
and the dynamical quenching of $\Delta\Omega$ and $\Omega$
thus increase the period. Only a modest
decrease in $\Omega \Delta \Omega$ makes the dynamo number subcritical,
after which the field decays.


The notches that occur in the
decay of the rotational quantities in Figs. 2-5 
occur during the phase of each cycle when
the field is well away from its peak in amplitude.
At these times, the damping of the differential rotation
and rotation is weak, and their rapid decay is arrested.
Note that the Poynting flux drops below zero
during part of the field oscillations (see the top left panels
in Figs. 2-5), and so the rotational
quantities can even slightly increase during that time. 
This effect is small and for the most part the curves
represent decay to the values at which the dynamo number 
drops below the critical value for growth.

\section{Further Implications}

\subsection{Cycle Period as a Signature of Magnetic Influence}

If signatures of the time dependence of energy injection to SN could 
be constrained observationally, the
quasi-periodicity seen in the Poynting flux  would be a specific prediction 
of the influence on a large scale dynamo.
Similarly, were a Poynting-flux driving cycle to operate 
in GRB progenitors, one might consider this to provide a source of variability.
Even if the role of magnetic fields is to supply a corona as 
needed in Ramirez-Ruiz \& Socrates (2005) rather than for directly
driving an outflow, evidence for the cycle period could
still be present.
However, only when $\beta_t << \beta_p$ can the Poynting flux 
be a competitive sink of the free energy
compared to the turbulent dissipation. When the latter
is the primary sink, the SN explosion can still be strongly aided, 
as emphasized
in Thompson et al. 2005, but neither evidence for a cycle  
period nor significant anisotropy would necessarily be expected.


\subsection{Pulsar kicks}


If a significant source of energy for the SN is 
rotational energy  of the shear layer (or the NS),
then only a small fraction of it needs to be asymmetrically
extracted across the poles to produce pulsar kicks of $\sim 200$km/s.
To see this, note that the estimated kick
kinetic energy $E_k$ compared to that available 
in $E_{PF}$
The ratio for a $1.4 M_\odot$ neutron star is
\begin{equation}
K= {E_{k}\over E_{PF}} \simeq 0.01\left({v_k\over 200{\rm km/s}}\right)^2
\left(E_{PF}/E_{rot}\over 0.1\right)^{-1} 
\left(E_{mag}\over 10^{51}\right)^{-1}
\end{equation}
where the integrated Poynting flux 
$E_{PF}$ is scaled to typical values for or 15$M_\odot$ progenitor.
That $K<<1$ is favorable as it implies  a weakly
asymmetric bipolar outflow can drive the kick.
This is more efficient than appealing to 
the indirect role  magnetic fields would play 
in producing a neutrino-driven kick (Lai \& Qian 1998).

\subsection{No neutron loading, no magnetars, and no contradiction with the MRI}
Several  other consequences of the interface dynamo in the context of 
previous work on MHD and field generation in the SN engine deserve mention.
First, because the magnetic field in the interface dynamo model is maximized 
at the base of the convection zone ($r_c-L\le r \le r_c$) rather than 
at the NS radius ($r_{ns}<r_c-L$),
the outflow would be less loaded with neutron-rich material than in a model 
that amplifies the field 
more strongly at the surface of the NS (Meier et al.\ 1976).  
Too much neutron loading by jets emanating from too deep within the engine 
would produce $r$-process material in excess of that observed.
Second, because the interface dynamo is transient, once 
the rotational energy is extracted, the field strength 
falls well below $10^{15}$G as we have shown.
In this respect, the presence of $\sim 10^{15}$ G  fields 
in the supernova progenitor 
does not imply an overabundance of magnetars.  
There is therefore no contradiction between magnetically driven SN 
associated with pulsars and present surface fields $<<10^{15}$G.  
The strong magnetar surface fields would not be produced by the interface dynamo, but instead perhaps from a dynamo within the NS (Thompson \& Duncan 1993) or from a fossil 
field from an earlier stage that was amplified by flux freezing, and dynamically relaxed 
during NS formation (Blackman \& Field 2004; Braithwaite \& Spruit 2004).

The interface dynamo field does not extend beneath $r_c-L$ because 
there is insufficient convective turbulence between $r_{ns}$ and $r_c-L$ 
for the field to grow exponentially there.
In the absence of another instability such as the MRI,
the field would only grow linearly for $r<r_c-L$.
Linear growth may have difficulty competing with the buoyant loss of field from
this region. In contrast, at $r_c$, exponential growth in the interface dynamo is facilitated even in the absence of the MRI via the 
turbulent pumping of the poloidal magnetic field into the shear region (e.g. Tobias et al.\ 2001).

The interface dynamo does not 
exclude the possible operation the MRI for $r< r_c-L$.
The MRI can in principle produce either a small-scale or a large-scale 
magnetic field (as defined in section 1), 
but the latter only when the turbulence resulting 
from the MRI has some pseudoscalar helicity. It may be that some field amplification at or below $r_c-L$ comes from 
the MRI and that the pressure gradient resulting from this field plays a 
role in the SN driver (Akiyama et al.\ 2003).
However, it remains to be understood how effectively the MRI operates
from radial shear 
as a generator of turbulence  in  pressure-dominated, convectively 
stable regions of stellar interiors (Balbus \& Hawley 1994).  
Latitudinal rather than radial differential rotation may be  more 
more important for the MRI in these environments.

\section{Conclusions}

The free energy in rotation and differential rotation comprise an important
source of energy that can be tapped for aiding, if not driving SN explosions.
This energy can be drained into  magnetic fields or into
heat from turbulent dissipation and both can help
power a SN: The magnetic field can 
mediate a SN either via direct Poynting flux driven outflows, or via a
corona that leads to a non-thermal neutrino spectrum (Ramirez \& Socrates 2005). The heat from turbulent dissipation of shear  
can conspire with neutrino heating
to drive the  explosion (Thompson et al. 2005).
Here we have investigated the relative importance of  
these two energy sinks by developing  a  dynamical, 
 transient, $\alpha-\Omega$ interface large scale dynamo model 
and applying it 
to the proto-supernova engine structures presented in Akiyama et al.\ (2003).
Our interface dynamo, like that commonly thought to be operating in the sun,
is a large-scale helical dynamo in which the shear layer providing
the $\Omega$ effect lies 
beneath the turbulent convection supplying the helical $\alpha$ effect.
For core-collapse proto-supernovae engines, the convection is driven 
by shock heating 
and the shear layer beneath it derives 
from the initial stellar core collapse.
Unlike the sun however, the much stronger shear for the SN progenitor comes from the
initial collapse and is not necessarily reseeded by convection.
We have therefore incorporated 
the dynamical backreaction of the growing field
on the shear, the decay of shear due to turbulent dissipation, 
and the extraction of rotational energy via Poynting flux.
Although we also parameterize  $\alpha$ quenching and buoyancy,
the magnetic quenching and/or turbulent 
dissipation of the shear dominates the  dynamo quenching
and makes the dynamo transient.

The transient dynamo applied to a 15$M_\odot$ progenitor 
with maximal inner shear layer rotation rates of $200$rad/s  
lasts $\sim 10-50$ seconds, leading to 
large-scale toroidal fields of strength $5 \ts 10^{14}$G and large-scale 
poloidal fields of strength $\sim 10^{13}$G 
at the base of the convection zone. 
The poloidal magnetic field has a significantly lower magnitude 
than the toroidal field, lowering the peak Poynting flux 
(which depends on the product 
of the two field components) compared to previous estimates 
that invoke the magnitude of the magnetic energy 
(e.g. Wheeler, Meier, \& Wilson 2002).
However extracting even 10\% of the available rotational energy 
via  the integrated Poynting flux is influential in driving
or making the SN anisotropic (Wheeler et al. 2000)
because the binding energy above the core for a $15M_\odot$ progenitor
is  only $\sim 5\ts 10^{50}$erg (Woosley, Heger, Weaver 2002). This 
is less than the total rotational free energy  ($\sim 10^{51}$erg/s) 
for cores rotating with $\sim 160$ rad/sec.
In addition, extracting  a modest $1\%$  of the 
into pole-asymmetric outflows
 can supply observed pulsar kicks.
Faster rotation rates than invoked here can supply even more 
energy, but very high rotation rates are beyond
the applicability of the  
structure model we adopted from Akiyama et al. (2003),
%

A signature of the influence of a large scale 
magnetic dynamo is the $\sim 1$ sec 
cycle periods which are also reflected in the Poynting flux energy deposition
to the SN explosion (Figs. 2-5).
Despite the importance of magnetic fields, 
it may be that only a fraction (determined by how
far the toroidal field can diffuse downward in a cycle period) 
of the shear layer is available for magnetic energy 
amplification. As seen in Figs 2 and 3, the 
dominant energy sink for the integrated 
shear layer is heat. A SN induced from this heat
would likely be less anisotropic than in the predominantly 
magnetically driven case.


Our dynamo formalism 
differs from the $\alpha-\Omega$ dynamo 
of Duncan \& Thompson (1993) which operates deeper inside the NS.
%
Our results are also complementary to the MRI-based estimates of Akiyama et al.\ (2003), 
who obtained magnetic energies associated with a $10^{15}$ G field, but   
inferred these values from turbulent magnetic energy saturation 
of the MRI turbulence rather than from specific values of a large-scale
magnetic field.  Also unlike these previous works, 
we  include the dynamical evolution of the shear and rotation.

Similar dynamo and outflow studies  for faster rotating SN progenitors,
and for more massive ``failed'' SN 
(MacFadyen \& Woosley 1999; MacFadyen, Woosley, \& Heger 2001) are also desired.Poynting flux produced from within such engines is a 
leading candidate to power GRB.  The predicted variability on time 
scales $\sim 1$ s, 
based on the relative phase of toroidal and poloidal fields in our dynamo
cycle, suggests an analogous variability in Poynting-flux dominated GRB models.

More work is also needed to understand the specific
details of how the two sinks (Poynting vs. heat) for the 
free energy actually drive the explosion.
For example, 
a key issue for any transient dynamo outflow model
is determining the dynamical formation of
a magnetic corona where the magnetic field can dominate the dynamics,
and how the large scale field opens up there to mediate the outflows.
Improved 3-D versions of 
self-consistent numerical studies that include both the 
amplification of a very weak seed field and the outflow generation  
(Matt et al.\ 2004; Moiseenko, Bisnovatyi-Kogan, \& Ardeljan 2004) 
warrant development.
Future  work specifically on transient dynamo theory should also 
include self-consistent treatments of buoyancy losses (e.g.
Cline, Brummell, \& Cattaneo 2003)
radial and latitudinal spatial dependence of the system, 
latitudinal differential rotation, 
dynamical treatments of the backreaction incorporating magnetic helicity
conservation (e.g. Blackman \& Field 2002;  Brandenburg \& Sandin 2004).
Distinguishing the relative roles of the MRI, convection driven dynamos,
and possibly shear-current effect dynamos (Rogachevskii \& Kleeorin 2003)
also warrants further work.
The extent to which the MRI might require latitudinal, rather than radial shear
in the pressure supported interior of a star
is of particular interest.  The MRI in such environments
has  thus far been studied only in the linear regime (Balbus \& Hawley 1994).

\ni {\bf Acknowledgments:} 
EGB acknowledges support from 
NSF grant AST-0406799, NASA grant ATP04-0000-0016, and the KITP of UCSB, where this work was supported in part by NSF Grant PHY-9907949.
JTN acknowledges a Horton Fellowship from the Laboratory for Laser Energetics.  JHT was supported in part by a Senior Visiting Fellowship at the
Isaac Newton Institute for Mathematical Sciences, University of
Cambridge, under EPSRC Grant N09176.
Thanks to the referee for inducing our inclusion of the nonlinear shear evolution, and Lifin Wang for comments.

\eject
\clearpage
\begin{table}[t]
\centering
\begin{tabular}{ccccccccccccc}
 \hline
            Case &$\Delta\Omega_0$&  $\Omega_0$  & $P_{M,p}$ &$c_t$ & $c_p$ & $Q$ & $N_{D,0}$ & $\tau_0$ & $\delta/L$ &
$E_{PF}$ & $E_{PF}/E_{dis}$ \\ 
\tableline
  &$rad/s$ & $rad/s$ & & & & & $$ & $s$& & & \\  \hline
 1& $80$  &  $160$ & $.04$ & $0.002$ & $.05$ & $5.0$  & $6.95$ & $1.57$ & $0.42$ &$2.70\times{10^{49}}$
& $.10$
\\
2& $80$  &  $160$ & $.02$ & $0.001$ & $.05$ & $5.0$  & $9.83$ & $1.57$ & $0.30$& $1.97\times{10^{50}}$
& $.83$
\\
3& $80$  &  $160$ & $.02$ & $0.0005$ & $.025$ & $5.0$  & $19.67$ & $1.57$ & $0.21$& $2.16\times{10^{50}}$
& $.99$
\\
4&  $80$  &  $160$ & $.01$ & $0.0005$ & $.05$ & $5.0$  & $13.91$ & $1.57$ & $0.21$&
$1.98\times{10^{50}}$ & $.90$
\\
5&  $80$  &  $160$ & $.01$ & $0.0003$ & $.03$ & $5.0$  & $23.18$ & $1.57$ & $0.16$&
$1.63\times{10^{50}}$ & $.76$
\\

6&    $80$  &  $160$ & $.004$ & $0.0002$ & $.05$ & $5.0$ & $21.99$ & $1.57$ & $0.13$ &
$1.35\times{10^{50}}$ & $.63$
\\
  7& $126$  &  $180$ & $.02$ & $0.001$ & $.05$ & $5.0$  & $13.09$ & $1.18$ & $0.23$&
$2.67\times{10^{50}}$ & $.45$\\
   8& $126$  &  $180$ & $0.1$ & $0.003$ & $.03$ & $5.0$  & $9.75$ & $1.18$ & $0.39$&
$2.98\times{10^{50}}$ & $.47$
\\
   9& $160$  &  $200$ & $.0125$ & $0.0005$ & $.04$ & $5.0$ & $24.50$ & $0.99$ & $0.13$ &
$2.15\times{10^{49}}$ & $.23$
\\
 10& $160$  &  $200$ & $0.5$ & $0.005$ & $.01$ & $5.0$  & $15.51$ & $0.99$ & $0.43$&
$5.98\times{10^{50}}$ & $.59$\\
   11& $160$  &  $200$ & $0.4$ & $0.004$ & $.01$ & $10.0$ & $17.38$ & $0.99$ & $0.38$&
$5.43\times{10^{50}}$ & $.53$  \\    \hline
 12& $80$  &  $160$ & $.02$ & $0.001$ & $.05$ & $5.0$ & $9.83$ & $1.57$ & $1.00$  &
$3.14\times{10^{50}}$ & $1.37$\\
 
 13& $80$  &  $160$ & $.02$ & $0.0005$ & $.025$ & $5.0$ & $19.67$ & $1.57$ & $1.00$ &
$9.01\times{10^{50}}$ & $5.43$ \\
14& $80$  &  $160$ & $.01$ & $0.0005$ & $.05$ & $5.0$  & $13.97$ & $1.57$ & $1.00$&
$35.65\times{10^{50}}$ & $2.96$ \\
15& $80$  &  $160$ & $.01$ & $0.0003$ & $.03$ & $5.0$& $23.12$ & $1.57$ & $1.00$  &
$9.19\times{10^{50}}$ & $6.62$ \\

 16& $80$  &  $160$ & $.06$ & $0.0018$ & $.03$ & $5.0$  & $9.46$ & $1.57$ & $1.00$&
$3.34\times{10^{50}}$ & $1.38$ \\
     17& $80$  &  $160$ & $.03$ & $0.001$ & $.03$ & $5.0$  & $12.69$ & $1.57$ & $1.00$ &
$3.49\times{10^{50}}$ & $1.58$\\
18& $80$  &  $160$ & $.03$ & $0.001$ & $.03$ & $0.5$ & $12.69$ & $1.57$ & $1.00$  &
$9.08\times{10^{50}}$ & $5.27$\\
      19&  $126$  &  $180$ & $.02$ & $0.001$ & $.05$ & $5.0$ & $13.09$ & $1.18$ & $1.00$&
$7.36\times{10^{50}}$ & $1.35$  \\
       20&   $126$  &  $180$ & $0.1$ & $0.003$ & $.03$ & $5.0$  & $9.72$ & $1.18$ & $1.00$ &
$4.01\times{10^{50}}$ & $.64$\\
         21&   $160$  &  $200$ & $.125$ & $0.0005$ & $.04$ & $5.0$  & $24.5$ & $0.99$ & $1.00$&
$1.47\times{10^{51}}$ & $1.99$ \\
            22&  $160$  &  $200$ & $0.5$ & $0.005$ & $.01$ & $10.0$  & $15.55$ & $0.99$ & $1.00$&
$5.60\times{10^{50}}$ & $.55$ \\
              23&  $160$  &  $200$ & $0.4$ & $0.004$ & $.01$ & $10.0$  & $17.38$ & $0.99$ & $1.00$&
$9.21\times{10^{50}}$ & $.91$
\\
  \tableline
\end{tabular}
\caption{Various dynamo model results and initial conditions.
From left to right, the columns measure the initial differential
rotation, the initial rotation, toroidal magnetic Prandtl number from
(\ref{prandtl}), the toroidal and poloidal diffusion constants
from Eq. (\ref{diffparam}), the buoyancy parameter $Q$ from (\ref{buoy}), 
the total intergrated Poynting
flux,
the ratio of integrated Poyting flux to the shear energy dissipated
by turbulence, the initial cycle period and the fractional thickness
 of the shear
layer tapped by the toroidal magnetic energy amplification. 
The models above the middle horizontal line 
invokes (\ref{delta}) while those below
the line use 100\% of the shear layer.}
\label{bintable}
\end{table}

\eject
\clearpage
\begin{figure}
\plotone{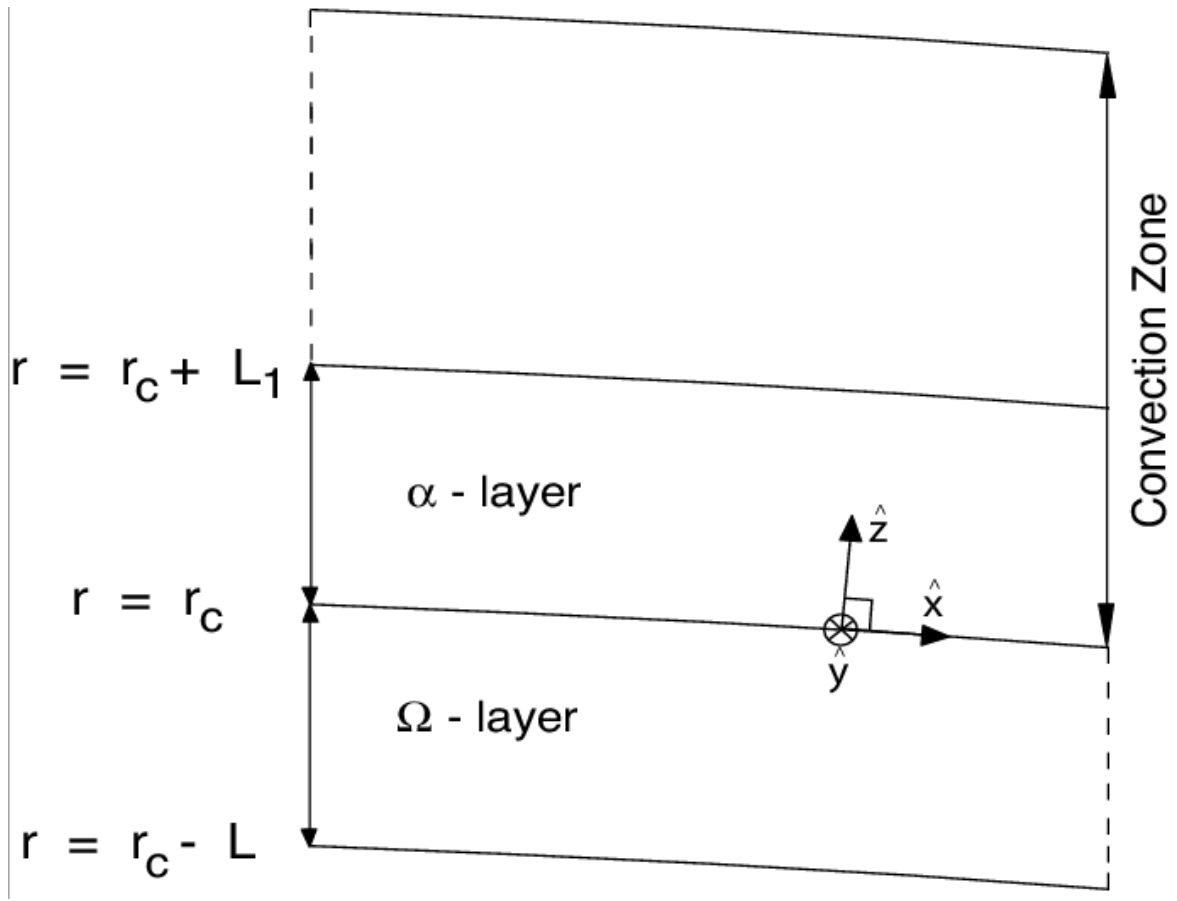}
\caption{Schematic meridional slice of the spherical dynamo engine.
The local Cartesian coordinate system is shown.
The $\alpha$-effect occurs in a layer of thickness 
$L_1$ equal to the local density scale height at the base 
of the convection zone ($r=r_c$).  The $\Omega$-effect occurs in a concentrated region of thickness $L$ which extends from the surface of the NS to the base of the convection zone.}
\end{figure}

\clearpage
\begin{figure}
\plotone{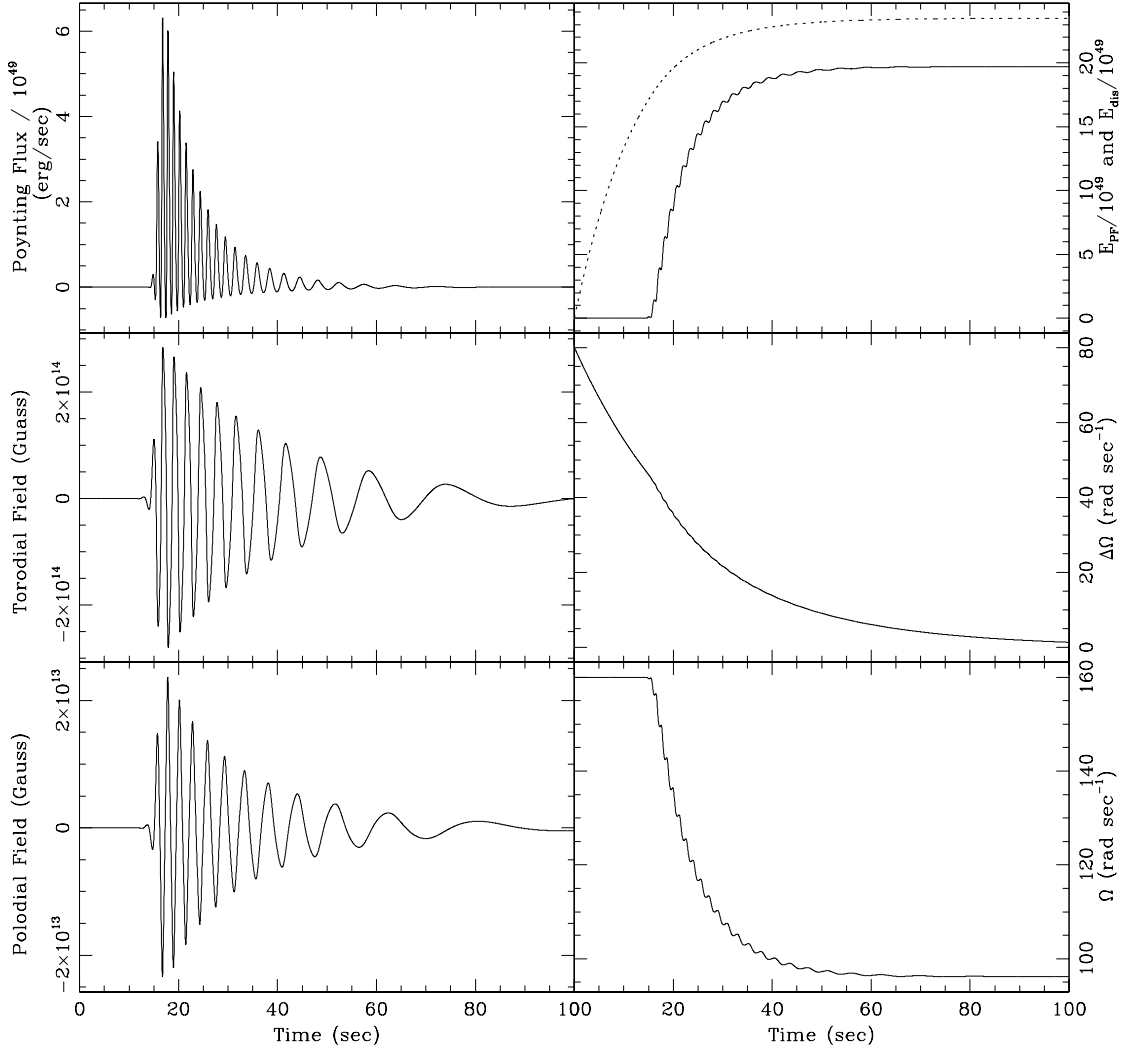}
\caption{
Toroidal and poloidal field strengths and Poynting flux are shown for $\Omega_0=160$ rad/sec and $\Delta\Omega_0=80$ rad/sec.  The solid line in the top right quadrant refers to the time integrated Poynting flux while the dotted line in the same graph shows energy lost to turbulent dissipation.  These solutions
correspond to row 7 of Table 1.
}
\end{figure}




\clearpage
\begin{figure}
\plotone{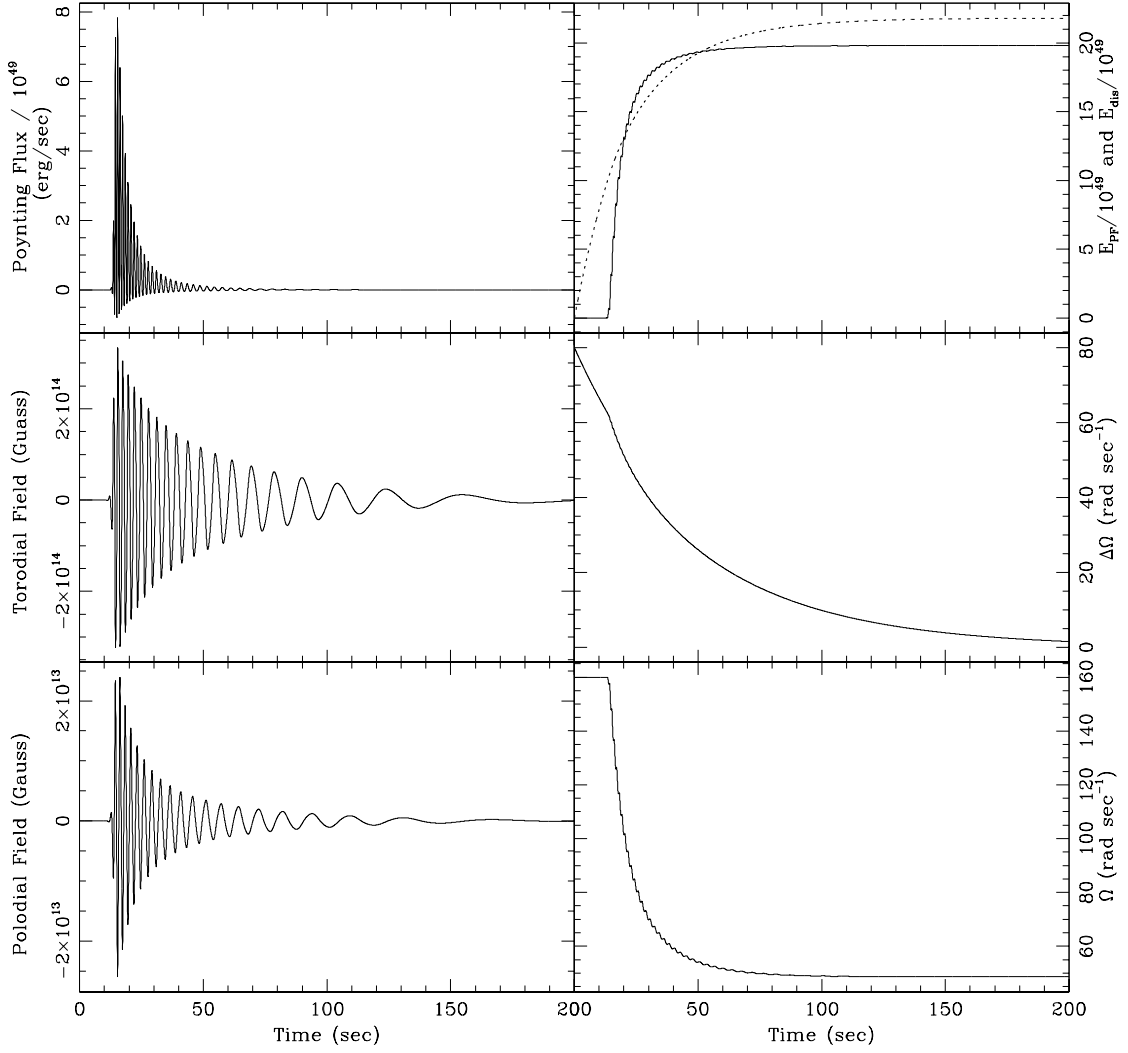}
\caption{Similar to Fig. 2 but for row 4 of Table 1.
As $\beta_t$ is lowered, the Prandtl number decreases which in turn increases $N_D$.  For higher dynamo numbers, the cycle period is lowered which results in torodial and polodial fields reaching their maximum value faster.}
\end{figure}


\clearpage
\begin{figure}
\plotone{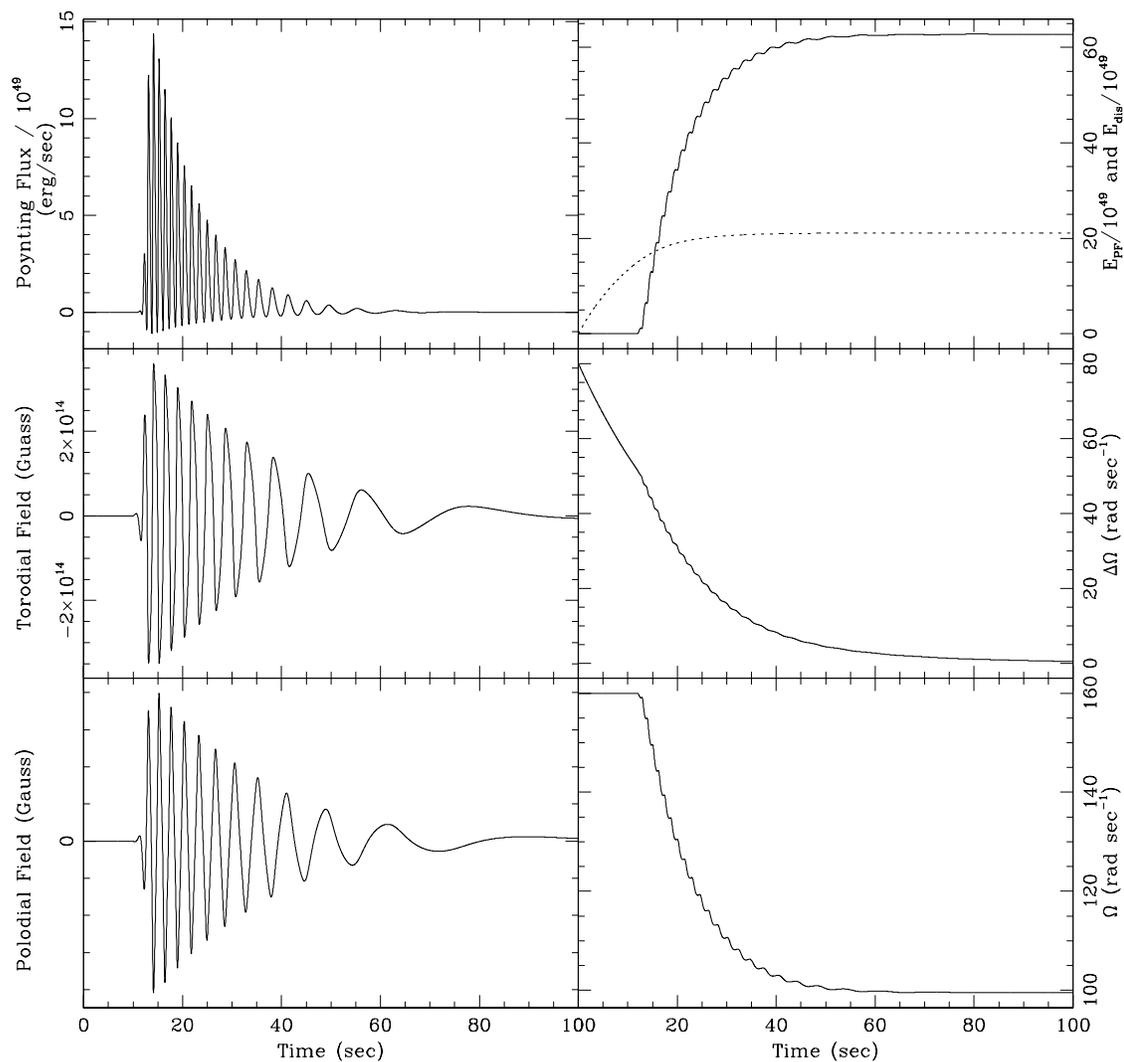}
\caption{Similar to Fig. 1 and 2 but with the parameters of
row 17 in Table 1. This  is 
an example of a model in which energy due to turbulent dissipation is large at early times after which magnetic energy dominates.  In this case, 100 percent of energy in the rotational layer is available for field growth.}  
\end{figure}

\clearpage
\begin{figure}
\plotone{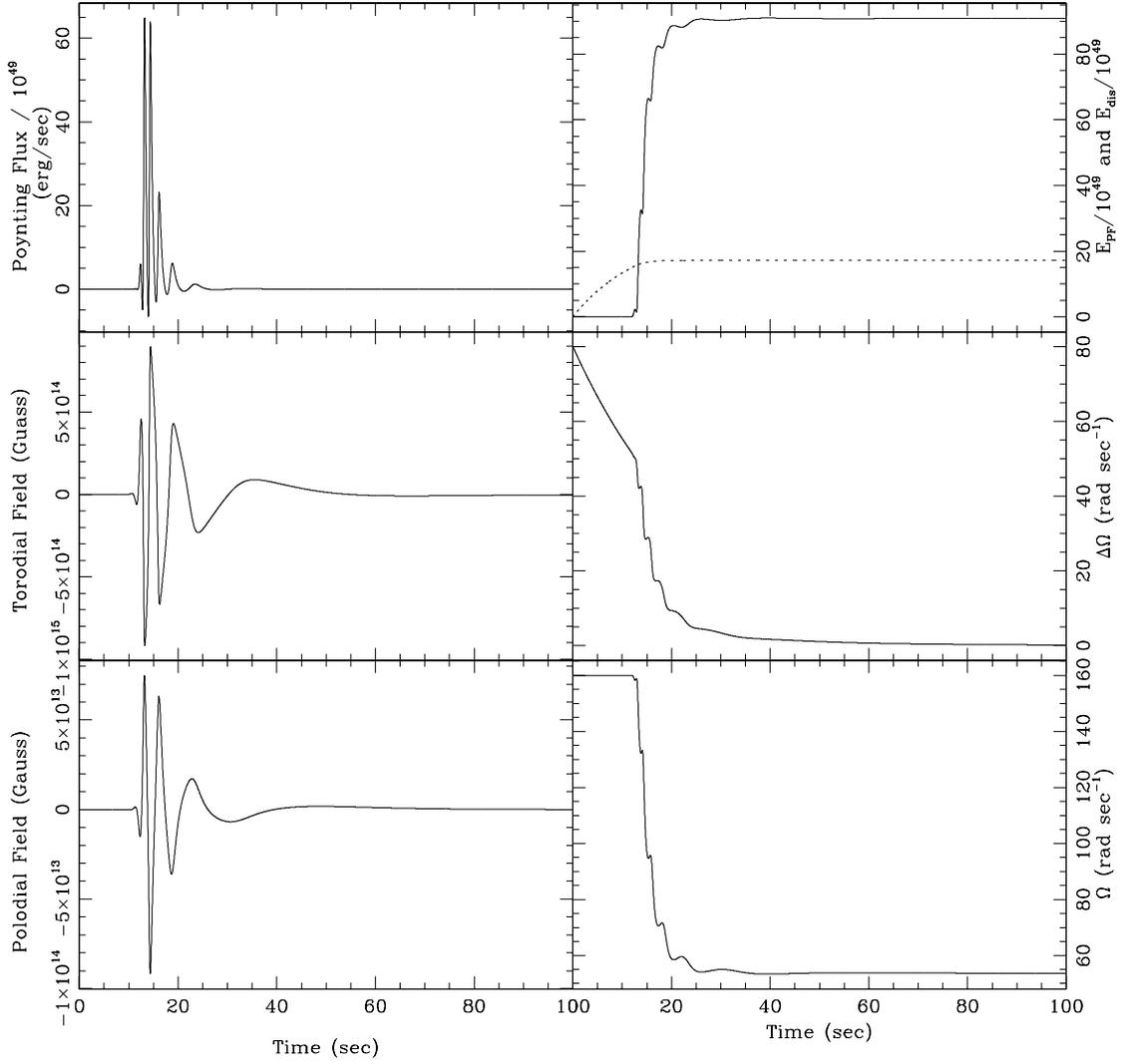}
\caption{Similar to  Fig. 4 but corresponding to row 18 in Table 1.  
In this case however, the coefficient for flux tube buoyancy is $Q=0.5$}
\end{figure}

\end{document}